# Visualisation of Law and Legal Process: An Opportunity Missed

Scott McLachlan[a,b,c,1], and Lisa C Webley[a]
[a] *Birmingham Law School, Birmingham University, Birmingham (UK)*
[b] *Risk and Information Management Group, Queen Mary University of London (UK)*
[c] *Health informatics and Knowledge Engineering Research Group (HiKER)*

**Abstract.** Visual representation of the law and legal process can aid in recall and discussion of complicated legal concepts, yet is a skill rarely taught in law schools. This work investigates the use of flowcharts and similar process-oriented diagrams in contemporary legal literature through a literature review and concept-based content analysis. Information visualisations (infovis) identified in the literature are classified into eleven described archetypal diagram types, and the results describe their usage quantitively by type, year, publication venue and legal domain. We found that the use of infovis in legal literature is extremely rare, identifying not more than ten articles in each calendar year. We also identified that the concept flow diagram is most commonly used, and that Unified Modelling Language (UML) is the most frequently applied representational approach. This work posits a number of serious questions for legal educators and practicing lawyers regarding how infovis in legal education and practice may improve access to justice, legal education and lay comprehension of complex legal frameworks and processes. It concludes by asking *how we can expect communities to understand and adhere to laws that have become so complex and verbose as to be incomprehensible even to many of those who are learned in the law*?

**Keywords.** data visualisation, legal process, legislation, flowcharts, lawmaps

## 1. Introduction

The law is a complex ecosystem, rife with ambiguity and discord such that it can be difficult to know how it works, which legislation should be applied, and in some cases whether a defendant's action falls within prohibited conduct [1-4]. Visual representation can aid those engaging with the law to organise, understand, improve collaboration and aid recall of complicated legal concepts has been understood for many years [5-7]. Legal visualisations can reduce confusion and miscomprehension for professional and lay-person alike [4], giving rise to a potential for preventing faulty decision-making and avoiding or mitigating errors and the significant costs associated with relitigating matters [8]. Legal education and practice are verbal in nature, and written text remains the primary presentation method common to both [4, 5]. Law students are rarely afforded the opportunity to develop the skills necessary to represent legal concepts using images or graphics, and in practice, it is more often the expert witness who presents visual artefacts to the court [4]. This work investigates the use of flowcharts and similar process-oriented diagrams to describe legislation or legal practice processes, seeking whether, how, and to what degree information visualisation techniques are being applied in contemporary legal literature and practice.

## 2. The Case for Information Visualisation

Information visualisation is the study of transforming data, information and knowledge into visual representations that can more easily convey meaning [9, 10]. Increasingly, it is recognised that effective information visualisation holds the key to unlocking access to, and understanding complexity in, data. In 1998 law graduate Mathew J McCloskey advocated that lawyers could benefit from learning how to *see* the law, and in doing so proposed an approach for *legal map-making* as a visualisation technique to improve comprehension of

---

the legal landscape [11]. While McCloskey promoted visualisation not as a graphic arts project, but as an approach to thinking about law, later authors have drawn only on examples more alike visual arts projects to support the misleading claim that *use of visualisation in the legal domain is growing* [12]. It has been observed that information visualisation can be applied to good effect for understanding complexity in legislation [13, 14], legal process and juridical deliberation [8]. However, the focus in legal literature has remained on *visual representation of law*, or how the culture of law is portrayed in images [12], film and television [15], rather than on *visualisation of law*, or approaches to diagrammatically represent legislation and case-law and demonstrate how they can be applied in a step-wise fashion. While, two decades after McCloskey's call to arms [4], we are not the first to identify that the legal domain still lacks effective and ubiquitous methods for information visualisation. This research is the first to support that contention with a review of the literature across that two-decade period.

Whether for professional or lay audiences, finding appropriate tools with which to visualise and communicate information is a challenging task [16, 17]. In medicine, studies have investigated the positive effect of visualisations on patients' and clinicians' understanding of medical risk [18-20]. However, outside of studies promoting their potential to improve legal research learning outcomes during jurist training [5], we were unable to identify any study that had evaluated the potential for visualisations to impact lawyers' and clients' understanding of either the law, or contemplations of litigation in criminal or civil matters.

Identifying the most appropriate way to visualise information and communicate its messages depends on the presenters' objective, the communication context, and the target audience [19, 21]. While target audience likes and dislikes should be taken into consideration, they must not be treated as the single defining standard [16]. For instance, some audiences prefer visualisations that are simpler, but simple graphs are not always able to convey complex information, which can lead to misunderstandings [22]. Research into the use of visualisation in the medical domain found that doctors performed worst with the visualisation format they liked best, and best with the one they most strongly disliked [23]. Effective data visualisation can mitigate issues that arise when deep insight is required to analyse data and make time-sensitive decisions [24]. Visualisations mitigate the complex issues of comprehension, interpretability and navigation as the target audience traverses large amounts of information [25]. Professionals recognise the effectiveness of visualisation techniques; however it is possible that professional scepticism regarding potential benefits acts as a significant limiting factor to their use [26]. While it is difficult to know whether the specific benefits identified in the visualisation literature are generalisable to the legal domain, the predominant benefits claimed include that the use of visualisation:

- allows the viewer to understand patterns and relationships not clearly visible within data [16, 24, 26];
- enhances communication of risk to a generic audience, especially one with low-numeracy skills [27];
- helps professionals to focus on, assimilate and recall issue-relevant aspects [24];
- improves problem solving and decision-making abilities [24, 27, 28].

In the medical domain the absence of or ineffective information visualisations has negative effects on clinical care, time efficiency and patient safety [29]. For law, the absence of research in this area makes it difficult to say what would constitute effective or ineffective visualisation, and therefore quantify what effect the absence of ubiquitous information visualisation may be having.

## 3. Method

A collection of literature databases of material written in English, including SCOPUS, DOAJ, AustLii, BAILii, HeinOnline and SSRN were searched using the term:

```
("law" or "legal") and
("process flow" or "process map" or "flowchart")
```

Content analysis [30, 31] was used to identify and record instances of concepts under investigation in the literature. General concepts were initially identified deductively based on the objectives of the review. These concepts were refined inductively on first reading of the literature. All data was collected using a structured excel spreadsheet from which graphs and other statistical data were developed.

## 4. Results

The literature search identified 574 articles for consideration. Literature not including some form of flow diagram, those not relevant to law or a legal process, and any articles where the included diagram was not in the context of a law or legal process were all rejected. As shown in Figure 1, this resulted in a collection of 71 articles for inclusion in this review.

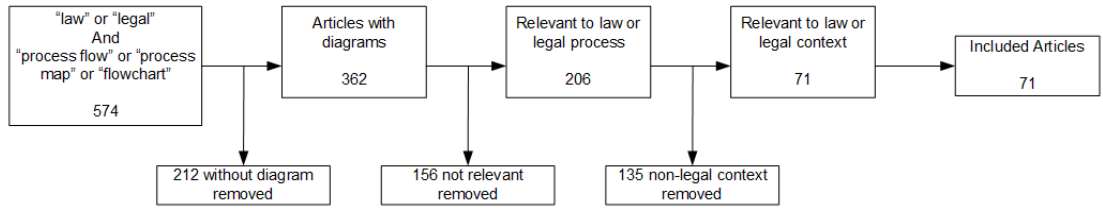

**Figure 1.** Literature search (PRISMA).

The concepts to be captured from the literature are shown in Figure 2. Just as it was possible for an article to contain one or more diagrams, it was also possible for an article to belong to one or more legal domains. For example: describing processes relevant to criminal law and procedure *and* criminal appeals [32]; dealing with issues for the remaining spouses and families of deceased military servicemen and women from the perspective of both medicolegal and family law [33]; or issues in forensic investigation and criminal law [34].

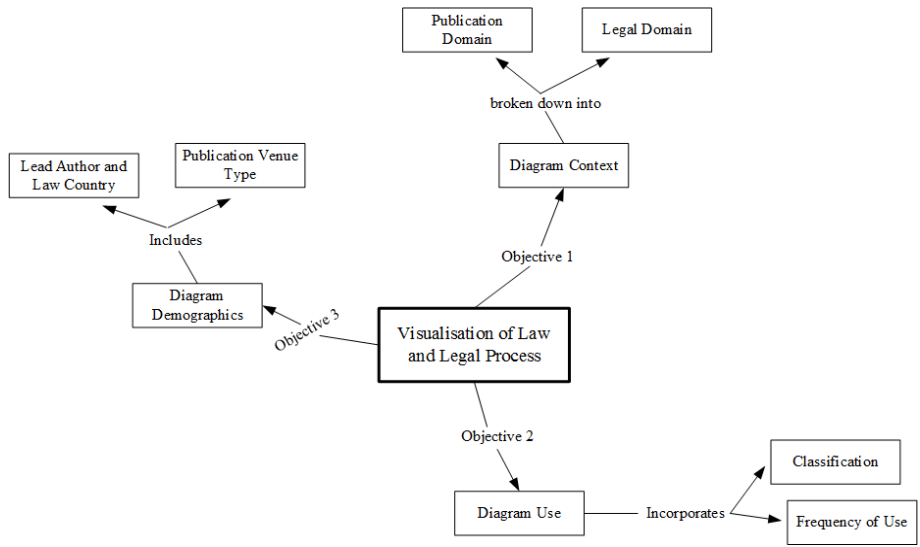

**Figure 2.** Concept Map for the Visualisation of Law and Legal Processes Literature Review

### 4.1. Classification of Diagrams

Diagrams in each paper were capable of classification into eleven archetypes which are described by order of frequency in Table 1, and by year of use and type in Figure 3.

Table 1: Information Visualisations in Legal Literature

| Visualisation Type | No. Found | Visual Representation | Description |
|---|---|---|---|
| Concept Flow | 49 | 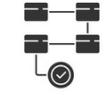 | Also known as *flowcharts* or *process maps*, this diagram visually describes the flow of work for a given activity, or the series of activities that result in a particular outcome. Examples included describing procedures for contesting matters in an Intellectual Property court [35], decision-making for international trade negotiations [36], the process for identifying express or implied restrictions in tenancies [13], and decision-making regarding state and federal tax aspects of perpetuity rules [37]. |
| UML Workflow | 15 | 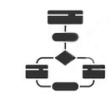 | Also known as an *activity diagram*. Provides a visual representation of the flow of work for a particular activity usually presented using the Universal Modelling Language (UML) notation framework. It describes activities and decisions, and in some cases the parties responsible for specific performance for each. Numerous examples were identified in the literature collection, including both simple [38] and complex [39] models that strictly adhered to the UML standard, along with many other analogous examples that loosely applied this standard [40, 41]. |
| Concept Map | 12 | 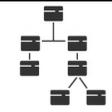 | This visualisation can take many forms and is capable of representing a broad range of relational information. Examples observed included tree-style maps to represent key aspects of individual litigated legal cases [42], the elements necessary to making a case [43], elements of financial contracts [44], and to prompt lawyers on enquiries that should be undertaken on behalf of the client to ensure due diligence in property matters [45]. |
| Process Map with Swim Lanes | 7 | 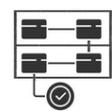 | A variation of the *concept flow* diagram where activities are described in lanes representing the responsible actor or source for that step. Examples included decision-making flows for multi-organisational prosecutorial investigations [34], sequences of responsibilities for different parties identified from legislation governing service contracts [46], and one extensive example identifying parties and rules from different legislative documents for the decision-making process governing requests for and release of medical records [14]. |
| Lifecycle Diagram | 5 | 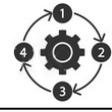 | Also known as a *cycle diagram*, this representation is used to show how a series of chronological events interact continuously, whether as a simple repeatable process or to incrementally improving practice. Cycles were observed for legal research [47], legal design [48], and general contract law [49]. |
| Mind Map | 4 | 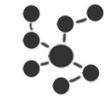 | The mind map is hierarchical and usually centres around a single target concept to show the relationships both between different sub-concepts, and between sub-concepts and the target concept, as an approach to visually organise information. Examples observed in the literature were generally educational tools to improve student comprehension and approaches to legal research and assignment writing [50]. |
| Timeline | 2 | 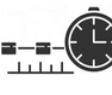 | The timeline presents a chronology of events or milestones that may be important to an undertaking or project. The timeline describes an overview of key points arranged along a line, usually from left to right, and doesn't generally stray into finer detail. Two examples were identified in the literature: a simple linear diagram describing the lifecycle duration of a contract [51], and a more comprehensive example describing the sequences for different events in the criminal justice system [52]. |
| UML Data Model | 2 | 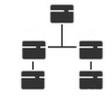 | This diagram is an object-oriented class model describing the overall structure of data or the complete relational structure of all tables and elements of a database. This diagram is usually presented using standard UML notation. Only two examples of this diagram type were identified in the literature collection [32, 53]. |

| Twist of Pearls | 1 | 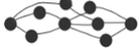 | The authors of the single paper [6] with this type of diagram describe it as symbolising the lifecycle of contracts. They portray the twist of pearls diagram as a visualisation where the twist, or string, represents the temporal continuum, and the pearls identify definition and evaluation points along the path to formation and implementation of a contract. A brief search was conducted, however no other description or example of this type of diagram could be located. |
|---|---|---|---|
| Relational Model | 1 | 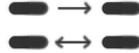 | Another novel type observed in only a single source [54]. The authors explain the diagram as one which is describing the relation between two contrasting items through time. Visually, this simplistic diagram presents as an arrow representing either the unidirectional or bidirectional relationship between two items being described. |
| Checklist | 1 | 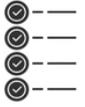 | Generally used by airline pilots and surgeons as a way to reduce failure by compensating for the limits of human memory and attention, at its most basic the checklist is a 'to do list' of necessary action items. One example was identified to prompt lawyers as to actions necessary in furtherance of dismissal of a matter [55]. |

*4.2. Further Analysis*

This research also sought to identify the existence of usage patterns for these diagrams, and if any identifiable patterns had changed over time or across legal domains. The frequency of diagram use by year of publication was recorded and is shown in Figure 3. As the most frequent type, it is not surprising that concept flows were found in every year where visualisations were identified. However, there were no discernible patterns in the distribution of visualisations during the two decades of literature included in this review.

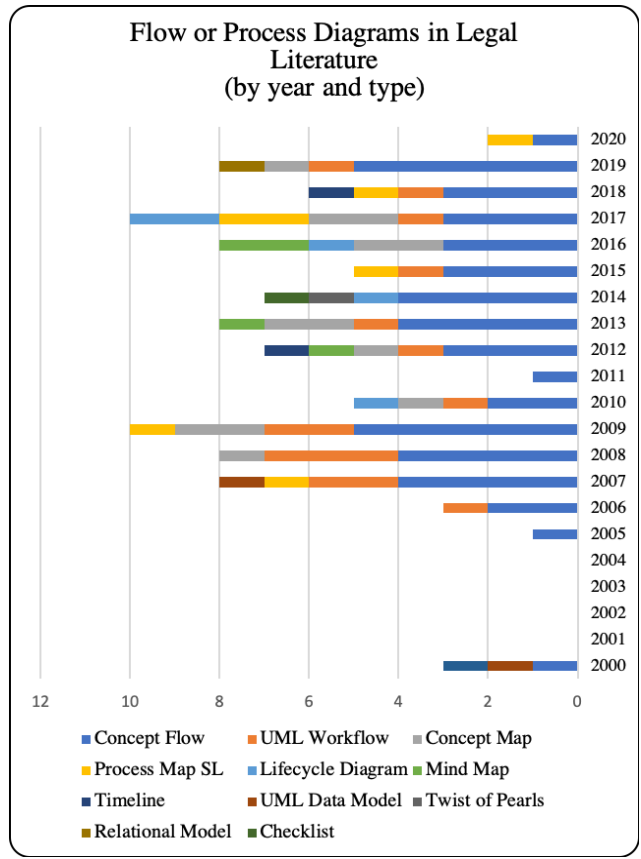

**Figure 3.** Flow or Process Diagrams in Legal Literature by Year and Type

The legal domain for both the *publication venue* and *subject matter context* were also recorded. Six publication venue domains were identified, with the majority of literature falling within the domains of *general law* or *legal education* as shown in Figure 4. Information Sciences was the only publication domain not to use the *concept flow* diagram type, relying instead on more scientific visualisations common to that domain, including those based on the UML notation.

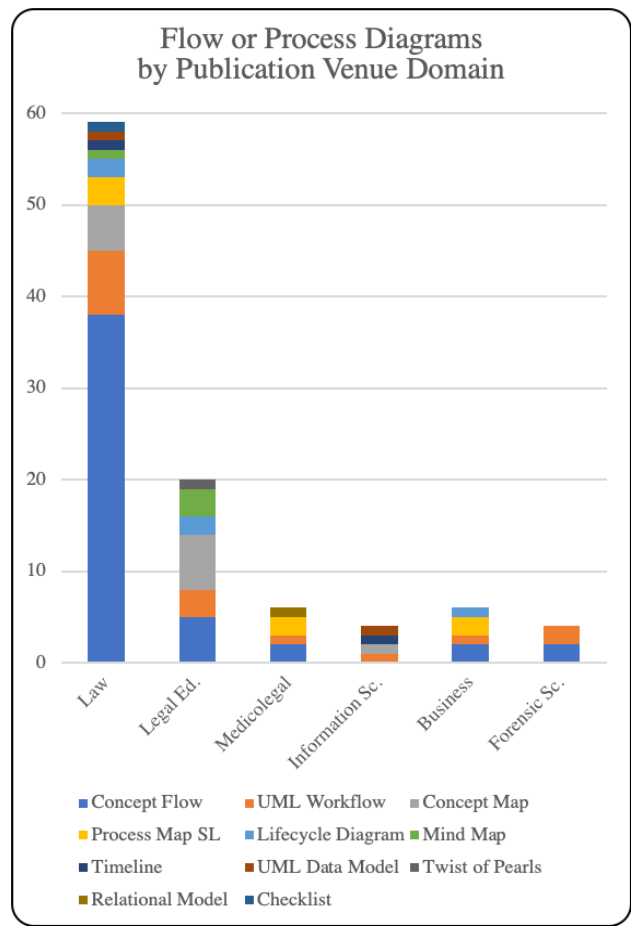

**Figure 4.** Flow or Process Diagrams by Publication Venue Domain

Twenty-six separate legal domains were identified from the subject context of the literature, as shown in Figure 5. The legal subject domain with the highest frequency of information visualisations was that of contract law. However, many of these could be characterised as focusing on smaller and often simpler sub-components of what are much larger processes or subject areas, such as identifying whether particular terms are consistent or cancel each other out in a contract [56], identifying of only the start (date of signing), notice period and end points of a contract on a timeline [51] shown in Figure 6, or the pathway process shown in Figure 7 for international money transfer of loan contract funds from a debtors bank to the creditor's bank [44].

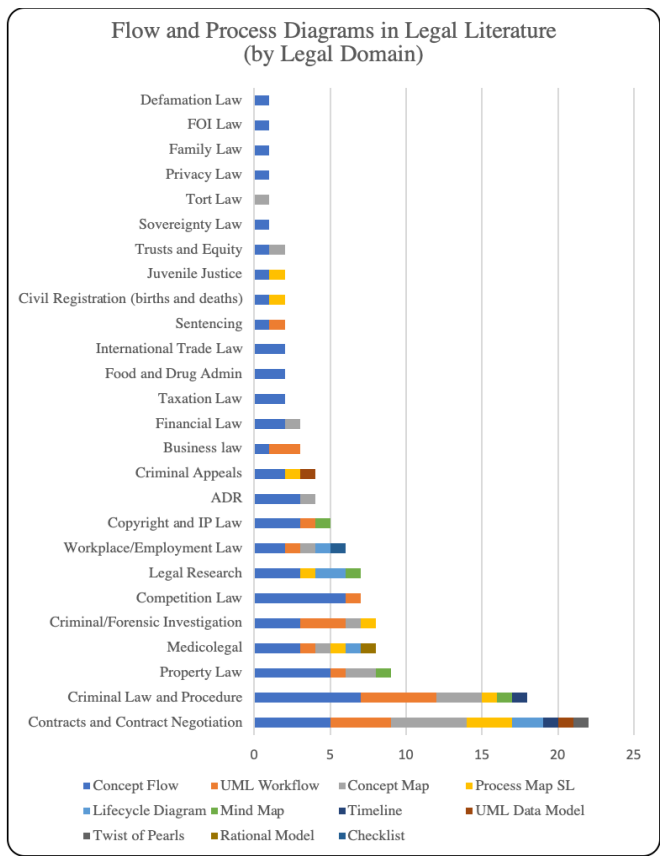

**Figure 5.** Flow or Process Diagrams by Legal Domain (subject context)

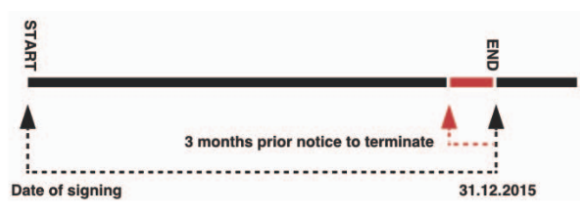

**Figure 6.** Contract Timeline from [51].

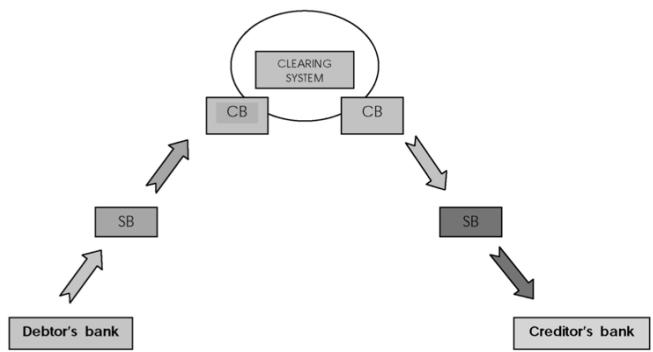

**Figure 7.** Loan Contract International Fund Transfer process from [44].

Diagrams in the area of Criminal Law and Procedure were observed almost as often, but appeared more mature. A key focus in this area was visualisation of critical decision-making processes such as those reproduced here for sentencing [57] in Figure 8, and civil jury deliberations [8] in Figure 9.

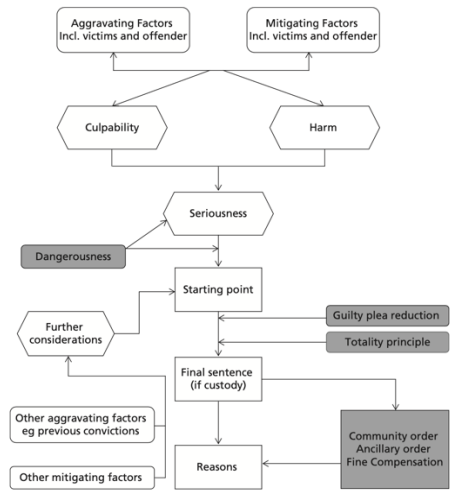

**Figure 8.** Criminal sentencing decision-making process from [57].

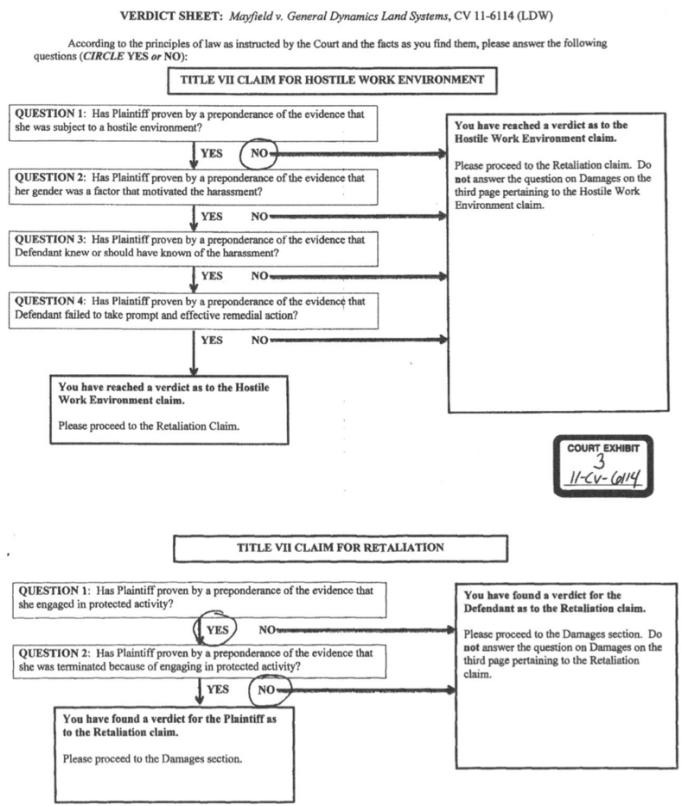

**Figure 9.** Civil jury deliberations decision-making process from [8].

*4.3. Diagrammatic representations of legislation*

Legislation is often imbued with latent pathways; where the application of one section leads to consequences in another, or where a rule may be reliant on consideration of matters described elsewhere, sometimes even in a different legislative instrument. Yet, in spite of this it was rare to find diagrams that visualised the reasoning or decision-making processes inherent to such legislation. Three examples were identified and are discussed here in order of complexity. The first were small swim lane tables showing relevant sections of one law governing the responsibilities of parties in a supplier/purchaser relationship [46]. The second provided a basic overview of the journey from registration of security to enforcement and seizure of financed property drawn from Australia's *Personal Property Securities Act* [50]. The third was a much more comprehensive swim lane flowchart reproduced here in Figure 10, of the various actors and New Zealand legislation that interact when a clinical records-holder makes decisions on whether to release an individual's personal medical record [14].

## 5. Discussion

Our literature review supports previous but quantitatively unsubstantiated claims that: (a) the largest collection of studies in this area of legal scholarship seek visualisation in contracts; and (b) that the most frequent approach for visually representing legal theory or process was the concept flow, or *flowchart* [46]. While it is well established that diagrams outperform text alone in supporting attention, information reasoning, comprehension and problem solving [46], our review shows that only a very small percentage of legal manuscripts present legal concepts in visual form. The remainder of this discussion considers four key areas, and in each it challenges those who draft, practice, research and teach law to consider how they could improve upon current approaches and increase professional and lay understanding of legal concepts and processes.

*Lay Comprehension*:
Much of the literature on lay-comprehension of law and legal concepts focuses specifically on the clarity and comprehensibility of judicial instructions, and how well juries understand and apply those directions when deliberating [58, 59]. While an important area for academic consideration, once a case has been handed over to the jury it is too late to begin contemplation as to whether those not trained in law but present in court have understood what occurred. Confusion around legal obligations and rights pervades all aspects of life. Improving the approach used to communicate law and legal concepts to lay people would increase participant agency, and improve legal practice and outcomes [60]. It has been shown that the average person has little comprehension of the content of most legislation, and their perspective on police, crime, judges, prisons and trials often does not exceed what is found in popular culture [61]. Indeed, it has even been observed that some of the documents prepared to advise the general public of their rights contain such complicated *legalese* that they become more incomprehensible than the legislation they describe [62].

In our role as lawyers, *can we find ways to more effectively explain to our clients the obligations, impacts and rights imposed by written documents and legislation?*

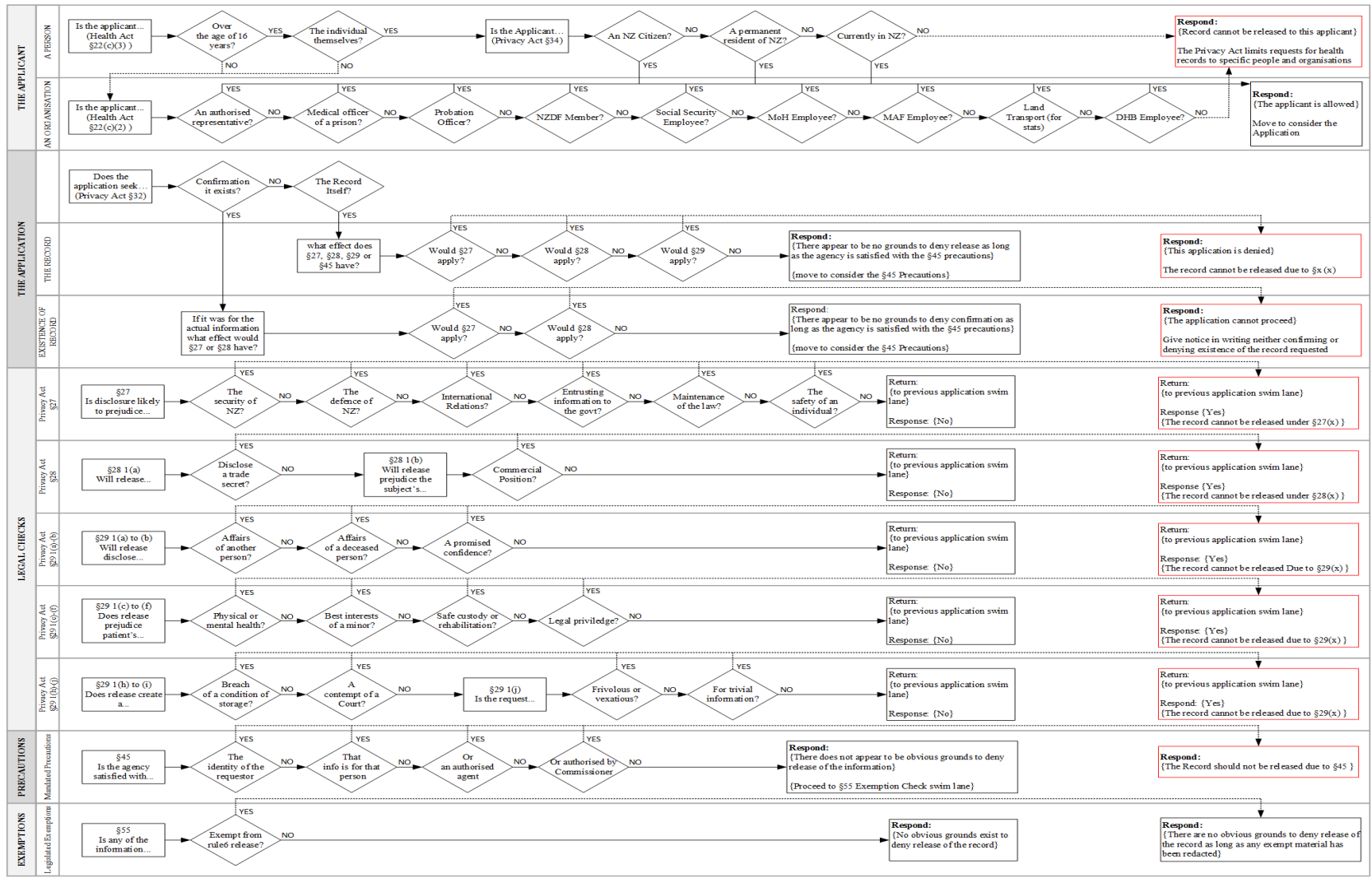

**Figure 10.** Release of Personal Medical records decision-making process from [14]

*Access to Justice*:

A small number of approaches have been employed to evaluate information visualisation of law and legal process with unrepresented litigants[2,3] [63, 64]. The overwhelming findings of these experiments has been to identify that visualisation, in combination with plain language explanation, increases overall comprehension and improves an untrained person's ability to raise valid legal arguments [65]. In this way, visualisation has been shown to have a positive effect on access to justice.

In our roles as legislators and court staff, *can we empower and enable everyone, especially those without legal training, to engage with and understand legislation and the application of law in everyday contexts?*

*Legal Education and Meaning*:

It has been said that lawyers possess a natural ability to create mental images of the law to situate themselves within the flow and circumstances of a case and visualise next steps based on precedent and past experience [11], yet even experienced lawyers can struggle to understand legislation[4]. While there is a strong call for information visualisation techniques to be taught [4, 50], aside from one reported example in 2011 at the University of Basel[5] there is little evidence in the curriculum and textbooks of most law schools that this call has been heard.

In our role as law school teachers, *can we more effectively inculcate law students with a broad appreciation for information visualisation as a tool to improve and enhance future legal practice*?

*Adoption*:

Australia began a push to improve the design of legislation in order to aid in comprehension of law through means other than text in 1995. The first call in Canada to make plain language and plain design legislation that would be visually inviting and more comprehensible came in 2000. Lawmakers in the United States received their first instruction to construct new law in *layman's terms* from President Nixon in 1972 [66]. An executive order so ignored that it was to be repeated by no less than three sitting presidents that followed: Carter, Clinton and Obama [66]. For the United Kingdom the call began in 2016, and to date there is little to suggest it has resulted in legislative structural change.

Legal researchers such as Margaret Hagan who runs the Visual Law project at the Open Law Lab of Stanford University[6] have undertaken research and spoken in favour of the need for visualising complicated legal text and concepts in *clear, digestible graphic presentations*[7]. Hagan's Visual Law project website provides a number of lay-approachable examples explaining caselaw and legal processes that were authored during the period from 2012-13. Many of these approaches fall somewhere within the remit of what has become known as the *plain language movement*, which has at times been derided and misunderstood while being equally lauded for its capacity for sense and clarity [67]. While the plain language movement is a significant step in the right direction, those advocating for legal information visualisation would argue plain text alone is insufficient; that members of the public with legal problems, as well as lawyers and law students, will always find plain text matched with visual tools and graphic alternatives easier to understand than text alone [4]. Aside from commentary from a former employee of a major Australian law firm describing his prior role as *Head of the Plain English Department,* where they developed plain language versions of service agreements for that firm's clients [67], there is little to suggest success for any of these *plain language* intentions for legislation and policy. Finally, while the authors of a 2012 review looking at visualisations of legislation sought to be positive in their conclusions about the prototypes they appraised and progress they considered had occurred in the two decades prior to their work, the outcome of their review [68], as ours, paints a bleak picture for the actual degree of adoption and impact of visualisation in law generally.

---

[2] https://lawhelpinteractive.org

[3] https://www.legalzoom.com

[4] This is why law, like medicine, comes inhabited by large number of specialties. And even within some specialties, like for example tax or intellectual property, there are sub-specialties where counsel only deal with very particular types of issues. The law is dynamic and constantly evolving - whether because new laws replace old or because caselaw changes the meaning or application of legislative provisions. Further, 'laws' are never settled or completely understood until they have been interpreted, tried and tested, in more than one case, by the courts.

[5] See "Producing, Analyzing, and Evaluation Legal Visualizations: A Pioneering Course at the Department of Law, University of Basel, Switzerland" (https://community.beck.de/gruppen/forum/producing-analyzing-and-evaluating-legal-visualizations-a-pioneering-course-at-the-department-of-law-unive)

[6] http://www.openlawlab.com/about/

[7] http://www.openlawlab.com/project-topics/illustrated-law-visualizations/

In our role as legal researchers, *can we encourage lawyers to adopt visualisation approaches into their legal processes and writings?*

## 6. Conclusion

One-fifth of the twenty-first century has already elapsed. However, even as computing technology, an avalanche of information, and increasingly more complicated new legislation continue to overwhelm both lawyer and lay-person, the call for plain-language laws coupled with information visualisation remains largely unanswered. The potential for well-crafted visualisations to improve law students, lawyers' and the general public's ability to engage with the law and legal system and to understand and contextualise both the law and legal processes cannot be overstated. When we consider the many thousands of academic articles, textbooks, case reports, websites, blog posts and other media published each year on an almost limitless range of legal topics, that the use of visualisations as observed in this literature review never exceeds ten in any given year is unfortunate. More than that, it demonstrates a collective failure to rethink and improve how we draft, teach, research, communicate and practice law to empower all in society. In exposing this continuing omission across the juridical sciences, this paper has also posed four challenges to those groups and posits one final question:

*How can we expect communities to be cognisant of and adherent to legislation that has become so verbose and complicated as to be incomprehensible, to even those who are legally educated?*